\newcommand\apjcls{1}
\newcommand\aastexcls{2}
\newcommand\othercls{3}

\newcommand\papercls{\aastexcls}
\documentclass[tighten, times, trackchanges, twocolumn]{aastex62}

\if\papercls \apjcls
\usepackage{apjfonts}
\else\if\papercls \othercls
\usepackage{epsfig}
\usepackage{margin}
\usepackage{times}
\fi\fi
\usepackage{ifthen}
\usepackage{natbib}
\usepackage{amssymb}
\usepackage[fleqn]{amsmath}
\usepackage{appendix}
\usepackage{etoolbox}
\usepackage[T1]{fontenc}
\usepackage{paralist}

\if\papercls \apjcls
\newcommand\aas{\ref@jnl{AAS Meeting Abstracts}}
\newcommand\dps{\ref@jnl{AAS/DPS Meeting Abstracts}}
\newcommand\maps{\ref@jnl{MAPS}}
\else\if\papercls \othercls
\usepackage{astjnlabbrev-jh}
\fi\fi

\if\papercls \aastexcls
\hypersetup{citecolor=blue, 
            linkcolor=blue, 
            menucolor=blue, 
            urlcolor=blue}  
\else
\usepackage[
bookmarks=true,           
bookmarksnumbered=true,   
colorlinks=true,          
citecolor=blue,           
linkcolor=blue,           
menucolor=blue,           
urlcolor=blue,            
linkbordercolor={0 0 1},  
pdfborder={0 0 1},
frenchlinks=true]{hyperref}
\fi

\if\papercls \othercls
\newcommand{\eprint}[1]{\href{http://arxiv.org/abs/#1}{#1}}
\else
\renewcommand{\eprint}[1]{\href{http://arxiv.org/abs/#1}{#1}}
\fi

\providecommand{\adsurl}[1]{\href{#1}{ADS}}

\makeatletter
\patchcmd{\NAT@citex}
  {\@citea\NAT@hyper@{%
     \NAT@nmfmt{\NAT@nm}%
     \hyper@natlinkbreak{\NAT@aysep\NAT@spacechar}{\@citeb\@extra@b@citeb}%
     \NAT@date}}
  {\@citea\NAT@nmfmt{\NAT@nm}%
   \NAT@aysep\NAT@spacechar\NAT@hyper@{\NAT@date}}{}{}

\patchcmd{\NAT@citex}
  {\@citea\NAT@hyper@{%
     \NAT@nmfmt{\NAT@nm}%
     \hyper@natlinkbreak{\NAT@spacechar\NAT@@open\if*#1*\else#1\NAT@spacechar\fi}%
       {\@citeb\@extra@b@citeb}%
     \NAT@date}}
  {\@citea\NAT@nmfmt{\NAT@nm}%
   \NAT@spacechar\NAT@@open\if*#1*\else#1\NAT@spacechar\fi\NAT@hyper@{\NAT@date}}
  {}{}
\makeatother

\makeatletter
\DeclareRobustCommand{\lowcase}[1]{\@lowcase#1\@nil}
\def\@lowcase#1\@nil{\if\relax#1\relax\else\MakeLowercase{#1}\fi}
\makeatother

\DeclareSymbolFont{UPM}{U}{eur}{m}{n}
\DeclareMathSymbol{\umu}{0}{UPM}{"16}
\let\oldumu=\umu
\renewcommand\umu{\ifmmode\oldumu\else\math{\oldumu}\fi}
\newcommand\micro{\umu}
\if\papercls \othercls

\else

\fi

\newcommand\microbar{\micro bar}

\let\oldsim=\sim
\renewcommand\sim{\ifmmode\oldsim\else\math{\oldsim}\fi}
\let\oldpm=\pm
\renewcommand\pm{\ifmmode\oldpm\else\math{\oldpm}\fi}
\newcommand\by{\ifmmode\times\else\math{\times}\fi}
\newcommand\ttt[1]{10\sp{#1}}
\newcommand\tttt[1]{\by\ttt{#1}}

\newbox{\wdbox}
\renewcommand\c{\setbox\wdbox=\hbox{,}\hspace{\wd\wdbox}}
\renewcommand\i{\setbox\wdbox=\hbox{i}\hspace{\wd\wdbox}}

\newcount\timect
\newcount\hourct
\newcount\minct
\newcommand\now{\timect=\time \divide\timect by 60
         \hourct=\timect \multiply\hourct by 60
         \minct=\time \advance\minct by -\hourct
         \number\timect:\ifnum \minct < 10 0\fi\number\minct}

\catcode`@=11

\newcommand\comment[1]{}

\newcommand\commenton{\catcode`\%=14}

\renewcommand\math[1]{$#1$}
\newcommand\mathshifton{\catcode`\$=3}

\let\atab=&
\newcommand\atabon{\catcode`\&=4}

\let\oldmsp=\sp
\let\oldmsb=\sb
\def\sp#1{\ifmmode
           \oldmsp{#1}%
         \else\strut\raise.85ex\hbox{\scriptsize #1}\fi}
\def\sb#1{\ifmmode
           \oldmsb{#1}%
         \else\strut\raise-.54ex\hbox{\scriptsize #1}\fi}
\newbox\@sp
\newbox\@sb
\def\sbp#1#2{\ifmmode%
           \oldmsb{#1}\oldmsp{#2}%
         \else
           \setbox\@sb=\hbox{\sb{#1}}%
           \setbox\@sp=\hbox{\sp{#2}}%
           \rlap{\copy\@sb}\copy\@sp
           \ifdim \wd\@sb >\wd\@sp
             \hskip -\wd\@sp \hskip \wd\@sb
           \fi
        \fi}
\def\msp#1{\ifmmode
           \oldmsp{#1}
         \else \math{\oldmsp{#1}}\fi}
\def\msb#1{\ifmmode
           \oldmsb{#1}
         \else \math{\oldmsb{#1}}\fi}

\def\supon{\catcode`\^=7}

\def\subon{\catcode`\_=8}

\def\supsubon{\supon \subon}

\newcommand\actcharon{\catcode`\~=13}

\newcommand\paramon{\catcode`\#=6}

\comment{And now to turn us totally on and off...}

\newcommand\reservedcharson{ \commenton  \mathshifton  \atabon  \supsubon 
                             \actcharon  \paramon}

\catcode`@=12
\reservedcharson

\if\papercls \apjcls

\else

\fi

\newcommand\chisq{\ifmmode{\chi\sp{2}}\else\math{\chi\sp{2}}\fi}
\newcommand\redchisq{\ifmmode{ \chi\sp{2}\sb{\rm red}}
                    \else\math{\chi\sp{2}\sb{\rm red}}\fi}
\newcommand\Teq{\ifmmode{T\sb{\rm eq}}\else$T$\sb{eq}\fi}
\newcommand\mjup{\ifmmode{M\sb{\rm Jup}}\else$M$\sb{Jup}\fi}
\newcommand\rjup{\ifmmode{R\sb{\rm Jup}}\else$R$\sb{Jup}\fi}
\newcommand\msun{\ifmmode{M\sb{\odot}}\else$M\sb{\odot}$\fi}
\newcommand\rsun{\ifmmode{R\sb{\odot}}\else$R\sb{\odot}$\fi}
\newcommand\mearth{\ifmmode{M\sb{\oplus}}\else$M\sb{\oplus}$\fi}
\newcommand\rearth{\ifmmode{R\sb{\oplus}}\else$R\sb{\oplus}$\fi}
\newcommand\molhyd{\ifmmode{{\rm H}\sb{2}}\else{H$\sb{2}$}\fi}
\newcommand\methane{\ifmmode{{\rm CH}\sb{4}}\else{CH$\sb{4}$}\fi}
\newcommand\water{\ifmmode{{\rm H}\sb{2}{\rm O}}\else{H$\sb{2}$O}\fi}
\newcommand\carbdiox{\ifmmode{{\rm CO}\sb{2}}\else{CO$\sb{2}$}\fi}
\newcommand\carbmono{\ifmmode{{\rm CO}}\else{CO}\fi}
\newcommand\ammonia{\ifmmode{{\rm NH}\sb{3}}\else{NH$\sb{3}$}\fi}
\newcommand\acetylene{\ifmmode{{\rm C}\sb{2}{\rm H}\sb{2}}
                        \else{C$\sb{2}$H$\sb{2}$}\fi}
\newcommand\ethylene{\ifmmode{{\rm C}\sb{2}{\rm H}\sb{4}}
                        \else{C$\sb{2}$H$\sb{4}$}\fi}
\newcommand\cyanide{\ifmmode{{\rm HCN}}\else{HCN}\fi}
\newcommand\nitrogen{\ifmmode{{\rm N}\sb{2}}\else{N$\sb{2}$}\fi}

\newcommand\nhyd{\ifmmode{N\sb{\molhyd}}\else{$N\sb{\molhyd}$}\fi}
\newcommand\nmethane{\ifmmode{N\sb{\methane}}\else{$N\sb{\methane}$}\fi}
\newcommand\nwater{\ifmmode{N\sb{\water}}\else{$N\sb{\water}$}\fi}
\newcommand\nmono{\ifmmode{N\sb{\carbmono}}\else{$N\sb{\carbmono}$}\fi}
\newcommand\ndiox{\ifmmode{N\sb{\carbdiox}}\else{$N\sb{\carbdiox}$}\fi}
\newcommand\nammonia{\ifmmode{N\sb{\ammonia}}\else{$N\sb{\ammonia}$}\fi}
\newcommand\nacetylene{\ifmmode{N\sb{\acetylene}}\else{$N\sb{\acetylene}$}\fi}
\newcommand\nethylene{\ifmmode{N\sb{\ethylene}}\else{$N\sb{\ethylene}$}\fi}
\newcommand\ncyanide{\ifmmode{N\sb{\cyanide}}\else{$N\sb{\cyanide}$}\fi}
\newcommand\nnit{\ifmmode{N\sb{\nitrogen}}\else{$N\sb{\nitrogen}$}\fi}

\newcommand\nhydrogen{\ifmmode{N\sb{\rm H}}\else{$N\sb{\rm H}$}\fi}
\newcommand\ncarbon{\ifmmode{N\sb{\rm C}}\else{$N\sb{\rm C}$}\fi}
\newcommand\nnitrogen{\ifmmode{N\sb{\rm N}}\else{$N\sb{\rm N}$}\fi}
\newcommand\noxygen{\ifmmode{N\sb{\rm O}}\else{$N\sb{\rm O}$}\fi}

\newcommand\pto{\ifmmode{p\sb{\rm to}}\else{$p\sb{\rm to}$}\fi}

\newcommand\rate{\textsc{rate}}
\newcommand\TEA{\textsc{TEA}}
\newcommand\sympy{\textsc{sympy}}
\newcommand\polyroots{\textsc{polyroots}}

\newcommand\eref[1]{(\ref{#1})}

\newcommand\HLT{HLT16}
\newcommand\HL{HL16}
\newcommand\HT{HT16}
\defcitealias{HengEtal2016apjAnalyticCO}{\HLT}
\defcitealias{HengLyons2016apjAnalyticHCO}{\HL}
\defcitealias{HengTsai2016apjAnalyticHCNO}{\HT}

\shorttitle{Reliable Analytic Thermochemical Equilibrium}
\shortauthors{Cubillos et al.}

\begin{document}

\title{Toward More Reliable Analytic Thermochemical-equilibrium Abundances}

\author{Patricio~E.~Cubillos}
\affiliation{Space Research Institute, Austrian Academy of Sciences,
              Schmiedlstrasse 6, A-8042, Graz, Austria}

\author{Jasmina~Blecic}
\affiliation{Department of Physics, New York University Abu Dhabi,
             PO Box 129188 Abu Dhabi, UAE.}

\author{Ian~Dobbs-Dixon}
\affiliation{Department of Physics, New York University Abu Dhabi,
             PO Box 129188 Abu Dhabi, UAE.}

\email{patricio.cubillos@oeaw.ac.at}

\begin{abstract}
  \hspace{0.0cm}\citet{HengTsai2016apjAnalyticHCNO} developed an
  analytic framework to obtain thermochemical-equilibrium abundances
  for {\water}, CO, {\carbdiox}, {\methane}, {\acetylene},
  {\ethylene}, {\cyanide}, {\ammonia}, and {\nitrogen} for a system
  with known temperature, pressure, and elemental abundances
  (hydrogen, carbon, nitrogen, and oxygen).  However, the
  implementation of their approach can become numerically unstable
  under certain circumstances, leading to inaccurate solutions (e.g.,
  ${\rm C/O} \ge 1$ atmospheres at low pressures).  Building up on
  their approach, we identified the conditions that prompt inaccurate
  solutions, and developed a new framework to avoid them, providing a
  reliable implementation for arbitrary values of temperature (200
  to $\sim$2000~K), pressure ($\ttt{-8}$ to $\ttt{3}$~bar),
  and CNO abundances ($\ttt{-3}$ to $\sim\!\ttt{2}\times$
  solar elemental abundances), for hydrogen-dominated atmospheres.
The accuracy of our analytic framework is better than 10\% for the more
abundant species that have mixing fractions larger than $\ttt{-10}$,
whereas the accuracy is better than 50\% for the less abundant species.
  Additionally, we added the equilibrium-abundance
calculation of atomic and molecular hydrogen into the system, and
explored the physical limitations of this approach.  Efficient and
reliable tools, such as this one, are highly valuable for atmospheric
Bayesian studies, which need to evaluate a large number of models.
We implemented our analytic framework into the {\rate} Python open-source
package, available at \href{https://github.com/pcubillos/rate}
{https://github.com/pcubillos/rate}.
\end{abstract}


\section{Introduction}
\label{introduction}

Understanding the physics governing planetary atmospheres is one of the
main goals of current and future research on transiting exoplanets.
Since the atmospheric temperature and composition are key properties
modulating the observed spectra of exoplanets, computing chemical
abundances is a fundamental step to link the observations to the
physical state of exoplanets.

Given the limited observational constraints currently existing for
exoplanet atmospheric composition, thermochemical equilibrium is the
educated guess of choice to estimate atmospheric abundances.
  We expect a medium to be in thermochemical equilibrium when it is
  sufficiently hot and dense, such that chemical reactions drive the
  composition faster than other processes.  This is the case for many
  sub-stellar objects and low-mass stars.  Even in the case of cooler
  atmospheres, where disequilibrium processes play a more important
  role, thermochemistry provides a starting point to contrast the
  impact of these other processes.  Consequently, thermochemistry has
  been widely studied to characterize the atmospheres of solar-system planets,
  exoplanets, and brown dwarfs
  \citep[e.g.,][]{Tsuji1973aaStellarAtmospheres,
    AllardHauschildt1995apjPHOENIX,
    AllardEtal1996apjlBrownDwarfAtmospheresGJ229B,
    FegleyLodders1996apjEquilibirumGJ229B,
    TsujiEtal1996aapBrownDwarfs,
    ZolotovFegley1998icarIo,
    BurrowsSharp1999apjBrownDwarfChemicalEquilib,
    LoddersFegley2002icarCNOsubstellarChemistry,
    VisscherEtal2006apjSPsubstellarChemistry,
    ZahnleEtal2009apjlSulfurPhotoch,
    VisscherEtal2010apjFeMgSiSubstellarChemistry,
    VisscherEtal2010icarJupiterChemistry,
    MosesEtal2011apjDissequilibriumHD209nHD189b,
    MarleyRobinson2015araaBrownDwarfs}.

Unfortunately, computing thermochemical-equilibrium abundances can
become a computationally demanding task, since it either requires one
to optimize a function with a large number of variables \citep[Gibbs
free energy minimization, e.g., ][]{BlecicEtal2016apsjTEA} or to solve
a large system of equations \citep[chemical reaction-rate networks,
e.g.,][]{TsaiEtal2017apjsVULCAN}.
Recently, \citet[][hereafter,
{\HLT}]{HengEtal2016apjAnalyticCO}, \citet[][hereafter,
{\HL}]{HengLyons2016apjAnalyticHCO}, and \citet[][hereafter,
{\HT}]{HengTsai2016apjAnalyticHCNO} developed an analytic formalism to
estimate thermochemical-equilibrium abundances for a simplified
chemical system composed of hydrogen, carbon, nitrogen, and oxygen.
They effectively turned the problem from a multi-variate optimization
into univariate polynomial root finding, which can be solved faster.
However, {\HT} detected numerical instabilities that prevent one from
using their method for arbitrary conditions.  Their proposed
stable---but simpler chemical network (without {\carbdiox},
{\acetylene}, nor {\ethylene})---does not appropriately account for
all plausible cases that might be encountered during a broad
exploration of the parameter space, for example, like those generated
in a Bayesian retrieval exploration.

In this article, we build upon the analytic approach of {\HT},
investigating the causes of the numerical instabilities and
identifying the regimes that prompt them.  We propose a new analytic
framework that considers multiple alternatives to compute the
equilibrium abundances and selects the most stable solution depending
on the atmospheric temperature, pressure, and elemental abundances.

In Section \ref{sec:preamble}, we lay out the theoretical preamble of
the problem to solve.  In Section \ref{sec:analytic}, we present our
variation of the analytic approach of {\HT}.  In
Section \ref{sec:fail}, we describe our improvements leading to more
reliable results.  In Section \ref{sec:implementation}, we present and
benchmark our open-source implementation of the analytic framework.
Finally, in Section \ref{sec:conclusions} we summarize our findings.

\section{Theoretical Preamble}
\label{sec:preamble}

For completeness, we reiterate the theoretical preamble for our
problem.  Consider a system composed of hydrogen, carbon, nitrogen,
and oxygen (HCNO), with known temperature ($T$), pressure ($p$), and
elemental abundances.  The problem is to determine the molecular
abundances for the system once it reaches thermochemical equilibrium.

{\HL} and {\HT} solve a simplified version of the network of reaction
rates (up to ten species and six reactions), which allows them to find
analytic expressions for the molecular abundances.  The six rate
equations they consider are:
\begin{eqnarray}
K\sb{1} & = & \frac{\nmono}         {\nmethane \nwater},    \label{eq:k1} \\
K\sb{2} & = & \frac{\nmono \nwater} {\ndiox},               \label{eq:k2} \\
K\sb{3} & = & \frac{\nacetylene}    {(\nmethane)\sp{2}},    \label{eq:k3} \\
K\sb{4} & = & \frac{\nacetylene}    {\nethylene},           \label{eq:k4} \\
K\sb{5} & = & \frac{\nnit}          {(\nammonia)\sp{2}},    \label{eq:k5} \\
K\sb{6} & = & \frac{\ncyanide}      {\nammonia \nmethane},  \label{eq:k6}
\end{eqnarray}
where $N\sb{X}$ are the normalized molecular number densities
normalized by the molecular hydrogen number density: $N\sb{X} \equiv
n\sb{X} / n\sb{\molhyd}$.  These values approximate the mole mixing
ratio of the species when molecular hydrogen solely dominates the
atmospheric composition.
The $K$ equilibrium coefficients are known values \citep[e.g.,
tabulated in NIST-JANAF thermochemical tables,][]{Chase1986jttJANAF},
which vary with temperature and pressure (see, e.g., {\HL} and {\HT}).

To complete the system of equations, {\HT} include the mass-balance
constraint equations for the metal elements:

\begin{eqnarray}
2\ncarbon     & = & {\nmethane} + \nmono + {\ndiox} + \ncyanide \nonumber \\
              &   & +\ 2\nacetylene + 2\nethylene,     \label{eq:c} \\
2{\nnitrogen} & = & 2\nnit + \nammonia + \ncyanide,    \label{eq:n} \\
2{\noxygen}   & = & {\nwater} + {\nmono} + 2{\ndiox}.  \label{eq:o}
\end{eqnarray}
For consistency, we have adopted the same nomenclature as {\HT}, where
the left-hand side terms are the elemental abundances normalized by
the hydrogen elemental abundance (in contrast to the molecular
equilibrium species on the right-hand side terms).  Hence, there is a
need for the correction factor on the left-hand side terms:
\begin{equation}
\frac{n\sb{\rm X}}{n\sb{\molhyd}}
          =  \frac{n\sb{\rm H}}{n\sb{\molhyd}} \frac{n\sb{\rm X}}{n\sb{\rm H}}
     \equiv  2 N\sb{\rm X}.
\end{equation}

Equations \eref{eq:c}--\eref{eq:o} assume that the elemental fraction
of metals (carbon, nitrogen, and oxygen) is negligible compared to
that of hydrogen; that all available hydrogen forms molecular hydrogen
(hence, ${n\sb{\rm H}}=2{n\sb{\molhyd}}$); and that the equilibrium
atomic abundances are negligible compared to the molecular abundances.
These conditions define the range in the parameter space where this
approach is valid.

\section{Analytic Equilibrium Abundances Revisited}
\label{sec:analytic}

One wants to solve a system of non-linear
Equations \eref{eq:k1}--\eref{eq:o}, where the molecular abundances
are the unknown variables.  The approach of {\HT} is to combine these
equations to create a univariate polynomial expression for one of the
molecules.  Then, one of the roots of such a polynomial corresponds to
the abundance of the molecule.

As already pointed out by {\HT}, a problem of this procedure is that
the approach is prone to numerical instabilities, which may lead to
inaccurate or even unphysical solutions.  We tested the analytic
implementation of {\HT} available in
VULCAN\footnote{\href{http://github.com/exoclime/VULCAN}
{http://github.com/exoclime/VULCAN}} (as of 2018 October).  This
implementation solves a reduced system of equations, neglecting
{\carbdiox}, {\acetylene}, and {\ethylene}, aiming to find a more
stable numerical solution; however, we find that even under this
simplification the code returns numerically unstable solutions (e.g.,
when {\ncarbon} > {\noxygen} at low pressures).  By unstable,
we mean chaotic behavior, where a solution varies significantly given
a small perturbation in the inputs.  These variations produce
inaccurate solutions, which may be as small as a few percent from the
expected value, or as severe as completely unphysical values.
Furthermore, this simplification limits the range of validity of the
code (e.g., this is not valid when ${\ncarbon} \gg {\noxygen}$,
because {\acetylene} and {\ethylene} can become the main
carbon-bearing species).  We thus seek alternative paths to construct
a polynomial expression.

\subsection{Hydrogen Chemistry}
\label{sec:hydrogen}

First, we adopt the more realistic assumption that equilibrium
hydrogen can form both molecular and atomic hydrogen.  Accordingly,
the hydrogen mass-balance equation is:
\begin{equation}
n\sb{\rm H} = 2n\sb{\molhyd} + n\sb{\rm H}\sp{\rm atom},
\label{eq:h}
\end{equation}
note that $n\sb{\rm H}$ is the total available amount of hydrogen,
whereas $n\sb{\rm H}\sp{\rm atom}$ is the amount that ends up as
atomic hydrogen under thermochemical equilibrium.  Then, following
{\HLT} for example, we compute the equilibrium constant between atomic
and molecular hydrogen:
\begin{equation}
K\sb{0} = \frac{(n\sb{\rm H}\sp{\rm atom})\sp{2}}{n\sb{\molhyd}}, \label{eq:k0}
\end{equation}
which we combine with Eq.~\eref{eq:h} to obtain the ratio $f =
n\sb{\rm H}/n\sb{\molhyd}$.  Since we assume a hydrogen-dominated
atmosphere, we can solve Eqs.~\eref{eq:h} and \eref{eq:k0}
independently of the other equations.  This is particularly relevant
at high temperatures and low pressures, where molecular hydrogen
starts to dissociate into atomic hydrogen.

\subsection{Multiple Paths to Solve the Analytic Problem}
\label{sec:newframe}

Allowing for atomic hydrogen introduces a modification on the
left-hand side of Eqs.~\eref{eq:c}--\eref{eq:o}:
\begin{eqnarray}
f\ncarbon     & = & {\nmethane} + \nmono + {\ndiox} + \ncyanide \nonumber \\
              &   & +\ 2\nacetylene + 2\nethylene,     \label{eq:newC} \\
f{\nnitrogen} & = & 2\nnit + \nammonia + \ncyanide,    \label{eq:newN} \\
f{\noxygen}   & = & {\nwater} + {\nmono} + 2{\ndiox}.  \label{eq:newO}
\end{eqnarray}

To solve these equations, we consider two cases: setting either
{\nwater} or {\nmono} as the polynomial variable.  Later, depending on
the case, we will prefer one over the other (the reasons will be clear
in Section \ref{sec:regimes}).  To begin, we plug in Eq.~\eref{eq:k2}
into Eq.~\eref{eq:newO} to remove {\ndiox} from the equations, leaving
an expression for {\nmono} as a function of {\nwater}:
\begin{equation}
\nmono  = \frac{f\noxygen - \nwater}{1 + 2\nwater/K\sb{2}},
\label{eq:mono}
\end{equation}
or {\nwater} as a function of {\nmono}:
\begin{equation}
\nwater  = \frac{f\noxygen - \nmono}{1 + 2\nmono/K\sb{2}}.
\label{eq:water}
\end{equation}

Now, we sequentially use the rest of the equations to find an expression
for all other variables in terms of the polynomial variable, i.e., use
Eq.~(\ref{eq:k1}), Eq.~(\ref{eq:k2}), Eq.~(\ref{eq:k3}),
Eq.~(\ref{eq:k4}), Eq.~(\ref{eq:newC}), Eq.~(\ref{eq:k6}), and
Eq.~(\ref{eq:k5}) to obtain, respectively:
\begin{eqnarray}
{\nmethane}   & = & {\nmono}/{(K\sb{1} {\nwater})},    \label{eq:methane}\\
{\ndiox}      & = & {\nmono}{\nwater}/ K\sb{2} \\
{\nacetylene} & = & K\sb{3}\ ({\nmethane})\sp{2}, \\
{\nethylene}  & = & {\nacetylene}/{K\sb{4}},   \\
{\ncyanide}   & = & f\ncarbon - \nmethane - \nmono - 2\ndiox \nonumber \\
              &   & - 2\nacetylene - 2\nethylene,   \label{eq:cyanide} \\
{\nammonia}   & = & {\ncyanide} / (K\sb{6} \nmethane), \label{eq:ammonia} \\
{\nnit}       & = & K\sb{5}\ ({\nammonia})\sp{2}.
\end{eqnarray}

Finally, Eq.~(\ref{eq:newN}) give us the expression for the polynomial:
\begin{equation}
2\nnit + \nammonia + \ncyanide - f{\nnitrogen} = 0.
\label{eq:nitrogen}
\end{equation}

We use the {\sympy} Python package to handle the algebra, which turns
the tedious task of finding the polynomial coefficients into a trivial
task.  {\sympy} allow us to work directly with Equations
(\ref{eq:mono}) to (\ref{eq:nitrogen}) rather than manually doing the
calculations.  {\sympy} outputs the polynomial coefficients (either
for {\nwater} or {\nmono}) as algebraic expressions as function of
$K\sb{1}$--$K\sb{6}$, $f$, {\ncarbon}, {\nnitrogen}, and {\noxygen},
which we copy and paste into our Python code (we provide the scripts
to construct these polynomials in the compendium for this article).
Modifying the equations under the different approximations (e.g.,
neglecting terms) becomes an effortless endeavor.

\section{Toward Reliable Equilibrium Abundances}
\label{sec:fail}

We identified three key modifications that improve the reliability of
the analytic approach, which we describe in the following subsections.
We restrict this analysis to our domain of interest: the region of
exoplanet atmospheres probed by optical and infrared observations.
Therefore, we explore pressures between $\ttt{-8}$ and $\ttt{3}$~bar;
temperatures between 200 and 6000~K; carbon, nitrogen, and oxygen
elemental abundances between $\ttt{-3}$ and $\ttt{3}\times$ the solar
values; and overall metal elemental fractions less than 10\% (i.e.,
$\ncarbon + \nnitrogen + \noxygen < 0.1$).

\subsection{Root-finding Algorithm}
\label{sec:roots}

To find the polynomial roots, we use the Newton-Raphson
algorithm \citep[NR; Section 9.5.6
of][]{PressEtal2002NumericalRecipes} instead of {\polyroots}, which is
used in VULCAN.  This choice yields several advantages: first,
{\polyroots} is known to provide inaccurate solutions under certain
circumstances (see {\polyroots} documentation); second, NR restricts
its solutions to the real numbers; third, NR allows one to set the
convergence precision of the root; and fourth, NR is generally faster
than {\polyroots}.

Newton-Raphson is an iterative method that finds one polynomial root
at a time, starting from a given initial guess value.  Thus, it is
imperative to start from an appropriate value.  We consider the
mass-balance constraints to set the starting position and boundaries
for the root (e.g., $0 \leq \nwater \leq f\noxygen$ when solving the
polynomial for $\water$).  We found the best results if we started at a
guess value close to the maximum boundary.  Assuming that there exists
a solution within such boundaries, if NR does not find a physically
valid root, we iteratively resume the root-finding step, decreasing
the initial guess by a factor of 10 in each iteration.  We
adopt the standard practice for convergence criterion, terminating NR
when the relative change between two consecutive iterations is less
than $\ttt{-8}$.

\subsection{Numerically Stable Regimes}
\label{sec:regimes}

By studying the behavior of the abundances under thermochemical
equilibrium, we identified the regimes where the analytic solutions
are numerically stable.  The oxygen chemistry is the most relevant
aspect to consider.  From Equation \eref{eq:newO}, $\nwater$,
$\nmono$, and $\ndiox$ `compete' to take up all of the available
{\noxygen}.  For the range of parameters studied here, only either
{\water} or CO dominate the oxygen chemistry.  When {\water}
dominates, the numerator in Eq.~\eref{eq:mono} takes very small
values.  In this case, small variations of the {\nwater} solution (of
the order of the numerical precision or less) can produce widely
different results for the rest of the species.  It follows that the
whole set of analytic solutions becomes numerically unstable.
Likewise, when CO dominates the oxygen chemistry, Eq.~\eref{eq:water}
produces numerically unstable solutions.  The task is then to identify
under which conditions {\water} or CO dominate the oxygen chemistry,
and avoid the unstable set of equations.

\begin{figure}[tb]
\centering
\includegraphics[width=\linewidth, clip]{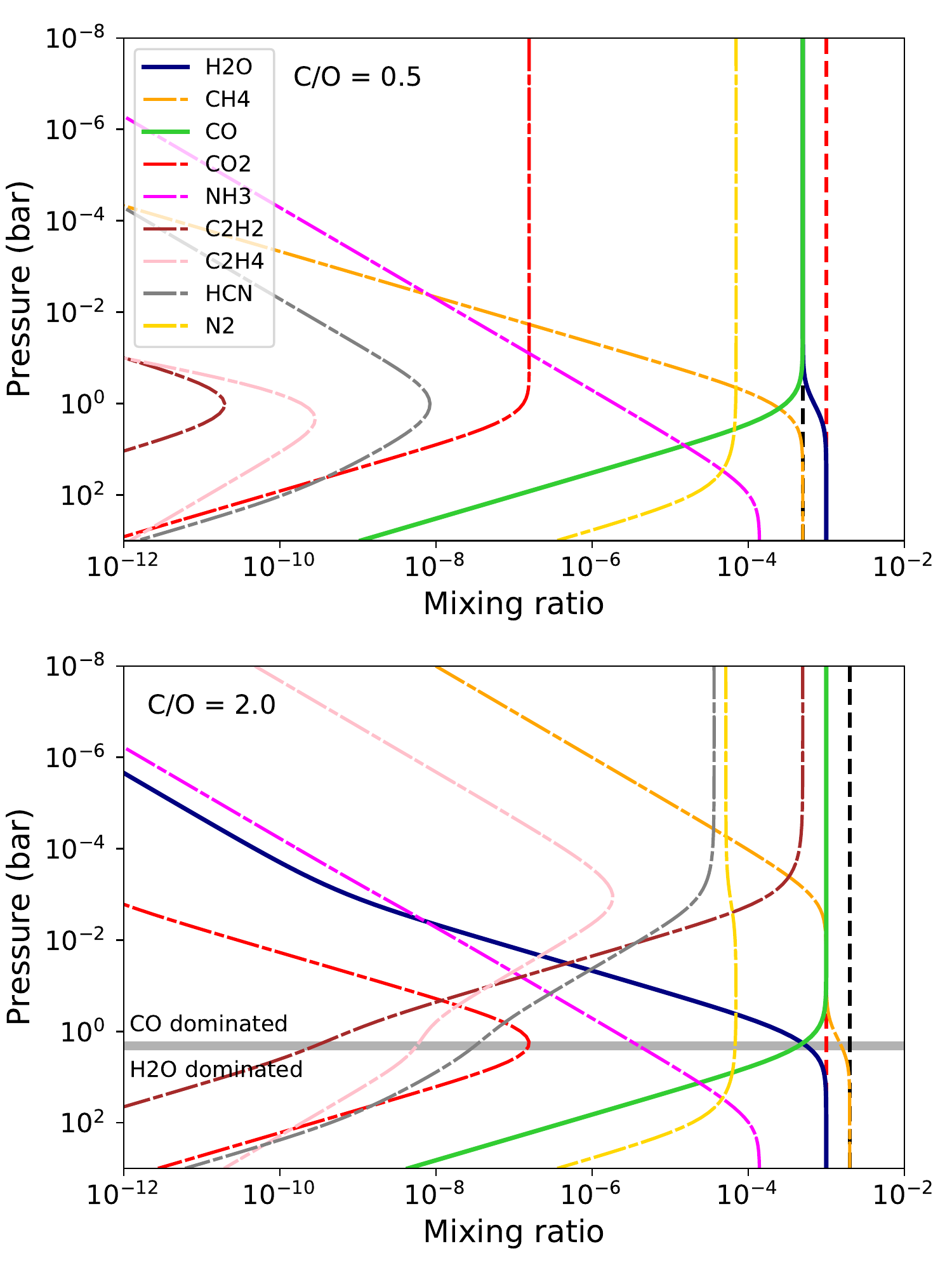}
\caption{
Thermochemical-equilibrium abundances for atmospheres with
$\ncarbon/\noxygen<1$ (top panel) and $\ncarbon/\noxygen>1$ (bottom
panel).  Both atmospheres have a fixed temperature of 1200~K,
$\noxygen=5\tttt{-4}$, and $\nnitrogen=7\tttt{-5}$. The black and red
dashed vertical lines denote $2\ncarbon$ and $2\noxygen$, which are
related to the maximum values that carbon- and oxygen-bearing species
can take, respectively.  The gray horizontal line in the bottom panel
denotes the turn-over pressure that separates CO- (above) and
{\water}-dominated (below) regimes.}
\label{fig:domains}
\end{figure}

Therefore, we consider two main regimes: carbon- and oxygen-dominated
atmospheres ($\ncarbon\geq\noxygen$ and $\noxygen > \ncarbon$,
respectively).  Oxygen-dominated atmospheres are simple to deal with;
since there is less carbon than oxygen available, $\nmono$ cannot
reach values equal to $f\noxygen$ (Figure \ref{fig:domains}, top
panel).  Therefore, we always adopt Eq.~\eref{eq:water}, and solve the
polynomial roots for CO.

Carbon-dominated atmospheres are more complex to deal with.  Here,
depending on the atmospheric properties, either $\nwater$ or $\nmono$
can reach values close to $f\noxygen$.  Fortunately, {\water} and CO
follow well-defined trends that allow us to estimate when either of
them dominates the oxygen chemistry.  In general, for a known
temperature and set of elemental abundances, we can find a turn-over
pressure (\pto) above which CO dominates, and below which {\water}
dominates (Figure \ref{fig:domains}, bottom panel).  To map how {\pto}
varies across our domain, we computed thermochemical-equilibrium
abundances over a four-dimensional grid of temperatures and C, N, and
O elemental abundances, using the open-source {\TEA}
package \citep{BlecicEtal2016apsjTEA}.  For each model, we then found
the pressure where CO and {\water} switch places as the dominant
species.  For our application, we model {\pto} as a fourth-order
polynomial in each of the four parameters, which matched the {\TEA}
{\pto} values at better than 20\% at $T < 3000$~K.  This accuracy is
sufficient for our purposes, since the range where an atmosphere
transitions between {\water} and CO is typically wider.  Then,
whenever we wish to compute the analytic abundances, we compare the
given pressure with {\pto} to determine whether we solve the
polynomial roots for {\water} (CO-dominated atmospheres) or for CO
({\water}-dominated atmospheres).

For the same reason, we avoid using Equation (\ref{eq:cyanide}) to
obtain $\ncyanide$, but rather combine Equations (\ref{eq:k5}) and
(\ref{eq:k6}) into Eq.~(\ref{eq:n}) to find a quadratic polynomial for
$\nammonia$:
\begin{equation}
2K\sb{5}\ (\nammonia)\sp{2} + (1+K\sb{6}\nmethane)\nammonia - f\nnitrogen = 0.
\label{eq:qnh3}
\end{equation}
Only one of those roots is positive; thus, selecting the right
solution is trivial.  Then we use Eq.~\eref{eq:ammonia} to obtain the
abundance of HCN.

\subsection{Proper Approximations}
\label{sec:approximation}

Although in Section \ref{sec:regimes} we developed a strategy to avoid
numerically unstable solutions, the root-finding algorithm may still
fail to return physically plausible abundances.  To ameliorate this
issue, we tested several approximations that lower the degree of the
polynomials, and hence, produce more robust results.  The challenge
is, given an atmosphere with an arbitrary temperature profile and
elemental abundances, determining under which regimes we can neglect
terms in Equations \eref{eq:mono}--\eref{eq:nitrogen} without breaking
the physics self consistency.

Initially, we considered three cases: neglecting all three
nitrogen-bearing species, producing sixth-order polynomials (which we
call the HCO solution); neglecting {\carbdiox}, {\acetylene}, and
{\ethylene}, producing sixth-order polynomials; and neglecting only
{\carbdiox}, producing eighth-order polynomials (HCNO solution).
After testing, we discarded the second option since it was the least
numerically reliable, and we found the other two approximations to be
sufficient to cover the parameter space.  Therefore, we proceed with
the sixth-order HCO and eighth-order HCNO polynomials.

In general, the simpler HCO polynomials produce more stable solutions
that are also faster to evaluate, making it our default choice.
Equations \eref{eq:c} and \eref{eq:o} help us to determine when the
HCO solution is appropriate (i.e., neglect the nitrogen-bearing
species).  Since the only link to nitrogen-bearing species relies on
HCN, when {\ncyanide} is negligible, the carbon and oxygen chemistry
becomes virtually decoupled from the nitrogen chemistry.  In this case,
we can solve the system of equations considering only the HCO
chemistry, without losing consistency for the entire system.

\begin{figure}[tb]
\centering
\includegraphics[width=\linewidth, clip,trim=0 110 30 20]{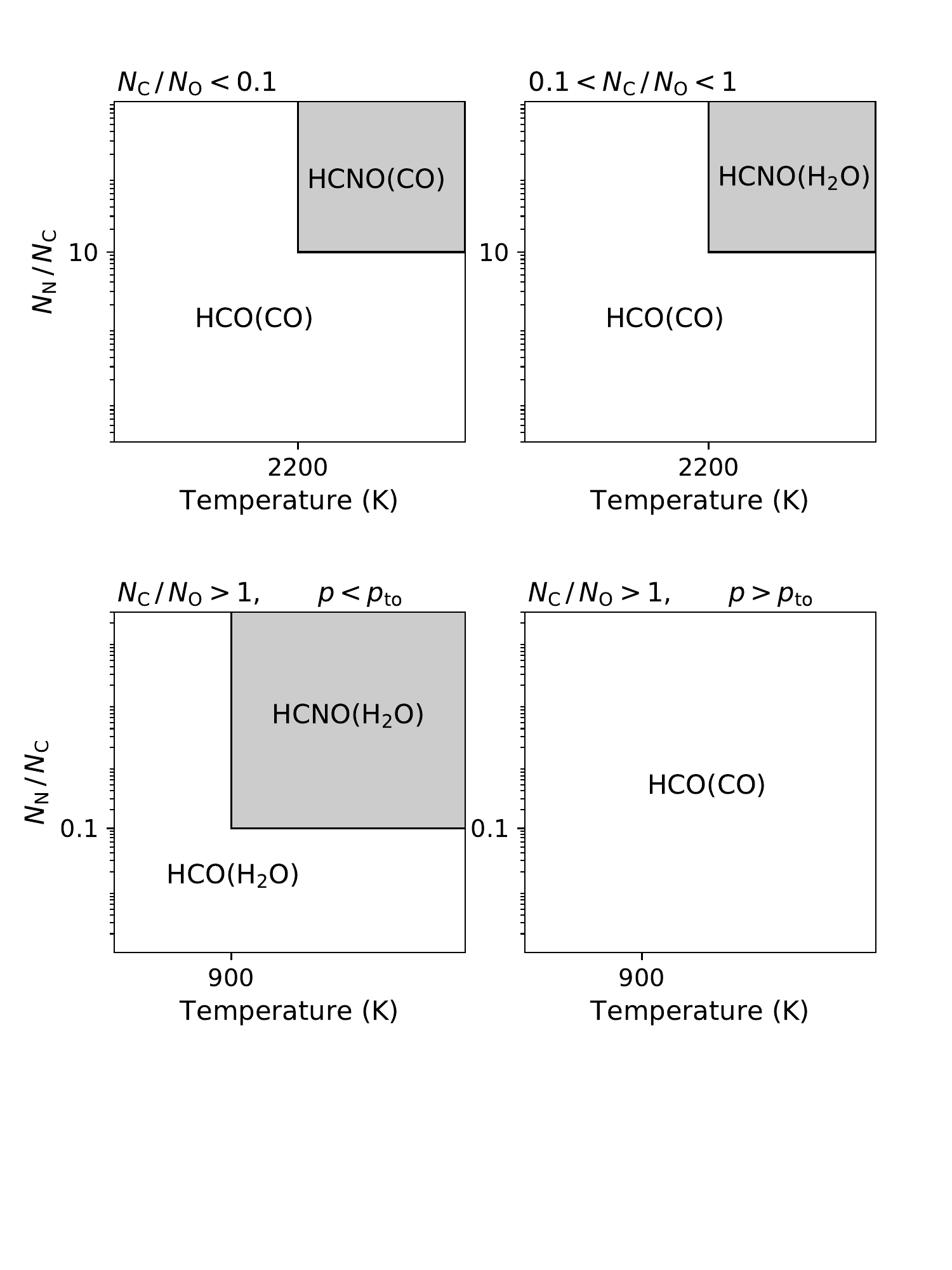}
\caption{
Schematic diagram of the solutions adopted for each case.  In general,
we adopt the HCNO polynomial at high temperatures and large
$\nnitrogen/\ncarbon$ ratio, because we need to consider the nitrogen
chemistry; otherwise, we adopt the HCO polynomial.  For
$\ncarbon/\noxygen<1$, we generally solve for the CO molecule because
it never dominates the oxygen chemistry; whereas for
$\ncarbon/\noxygen>1$, we use $\pto$ to chose the not-dominating
molecule between CO or {\water}.}
\label{fig:regimes}
\end{figure}

Overall, {\ncyanide} decreases with decreasing temperature, with
increasing pressure, or with decreasing {\nmono/\nwater}.  Therefore,
we set threshold values in $T$ and $\nnitrogen/\ncarbon$ to chose
between the HCO or the HCNO approach.  Similarly, we set threshold
values in $\ncarbon/\noxygen$ and use $\pto$ to chose solving the
polynomials for {\water} or CO.  We use this theoretical insight to
empirically find the threshold boundaries by evaluating all of our
solutions over the parameter space.  Figure \ref{fig:regimes} shows
which polynomial approximation and variable we adopt for each case.

For $\ncarbon/\noxygen < 1$ atmospheres, we find the best results by
applying the HCNO solution when $\nnitrogen/\ncarbon>10$ and
$T>2200$~K, for all other cases we apply the HCO solution.  As
expected, solving for CO produced more stable results, except for the
HCNO case when $0.1 < \ncarbon/\noxygen < 1$, where we solve for
{\water} instead.

For $\ncarbon/\noxygen \geq 1$ atmospheres, we find the best results
by applying the HCNO solution when $\nnitrogen/\ncarbon>0.1$ and
$T>900$~K, but only when $p<\pto$; for all other cases, we apply the
HCO solution.  We use {\pto} to decide to solve for the {\water}
(CO-dominated atmosphere) or CO polynomial ({\water}-dominated).

\section{Implementation and Benchmarking}
\label{sec:implementation}

We implemented the analytic framework described in this article into the
{\rate} open-source Python package, available
at \href{https://github.com/pcubillos/rate}
{https://github.com/pcubillos/rate}, which is compatible with Python
2.7 and 3.  The routine takes typically 5--10 ms to evaluate a
100-layer atmosphere on a 3.60~GHz Intel Core i7-4790 CPU.
We benchmarked the analytic abundances by comparing their results
against the {\TEA} code, computed under identical condition (i.e., same
temperature, pressure, elemental composition, and output species).

We focus this exploration on the range of atmospheric properties
probed by optical-to-infrared observations of sub-stellar objects.
Thus, we select a pressure range from 100 to $\ttt{-8}$~bar; a
temperature range from 200 to 6000~K; and metallicities from
$\ttt{-3}$ to $\ttt{3}$ times solar values.  Theoretical and
observational studies argue that sub-stellar objects may have C/O
ratios ranging from 0.1 to somewhat larger than unity (the solar C/O
ratio is 0.55), depending on the formation scenario and
evolution \citep[see, e.g.,][]{Madhusudhan2012apjCOratios,
MosesEtal2013apjCompositionalDiversity}.  Therefore, we further explore
C/O ratios from 0.1 to 5.0.

\begin{figure*}[p]
\centering
\includegraphics[width=\linewidth, clip]{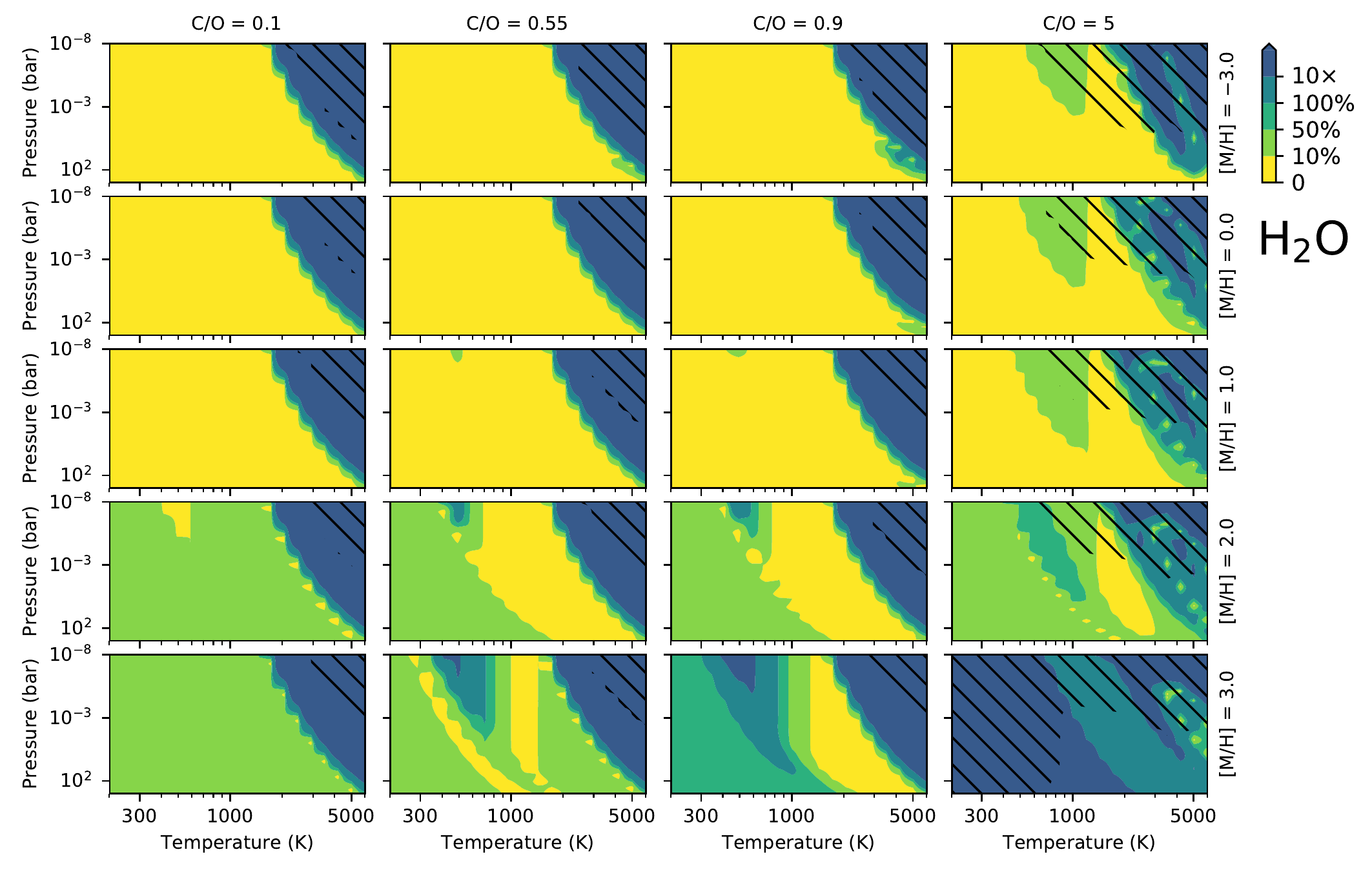}
\includegraphics[width=\linewidth, clip]{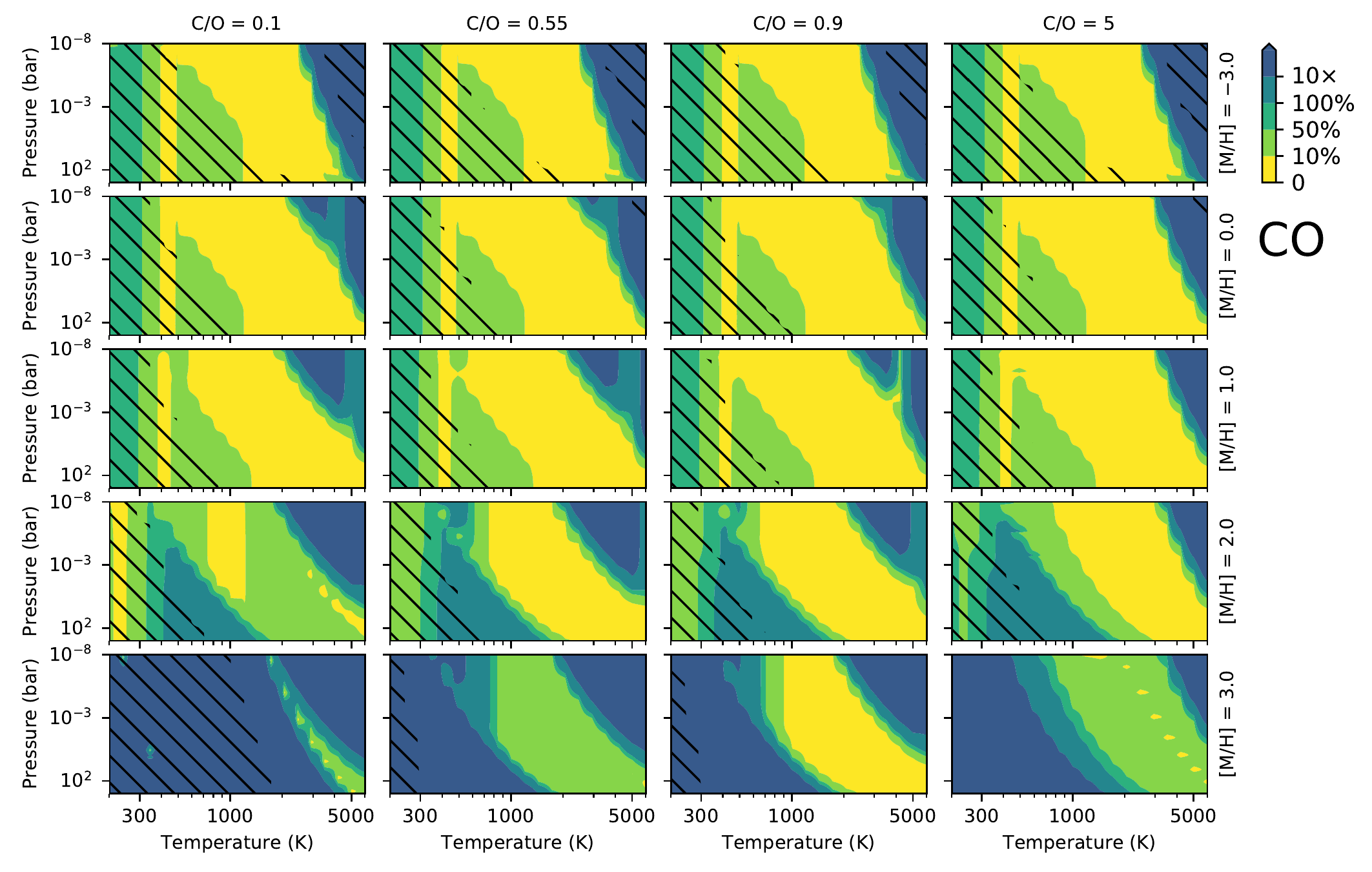}
\caption{
Difference between {\TEA} and {\rate} abundances for {\water} (top panels)
and CO (bottom panels) coded by color (see top-right color bar).
Each panels shows the {\TEA}--{\rate} difference in percentage as a function
of temperature and pressure.  For each species, the columns and rows
show different values of C/O ratios and metallicities (see labels at
the top and right side of the panels).  Note the non-linear scale in
metallicity (there are little differences for any sub-solar
metallicity run).  The hatched areas denote the regions where the
species abundance drops below a mixing fraction of $\ttt{-10}$.
}
\label{fig:bench1}
\end{figure*}

\begin{figure*}[p]
\centering
\includegraphics[width=\linewidth, clip]{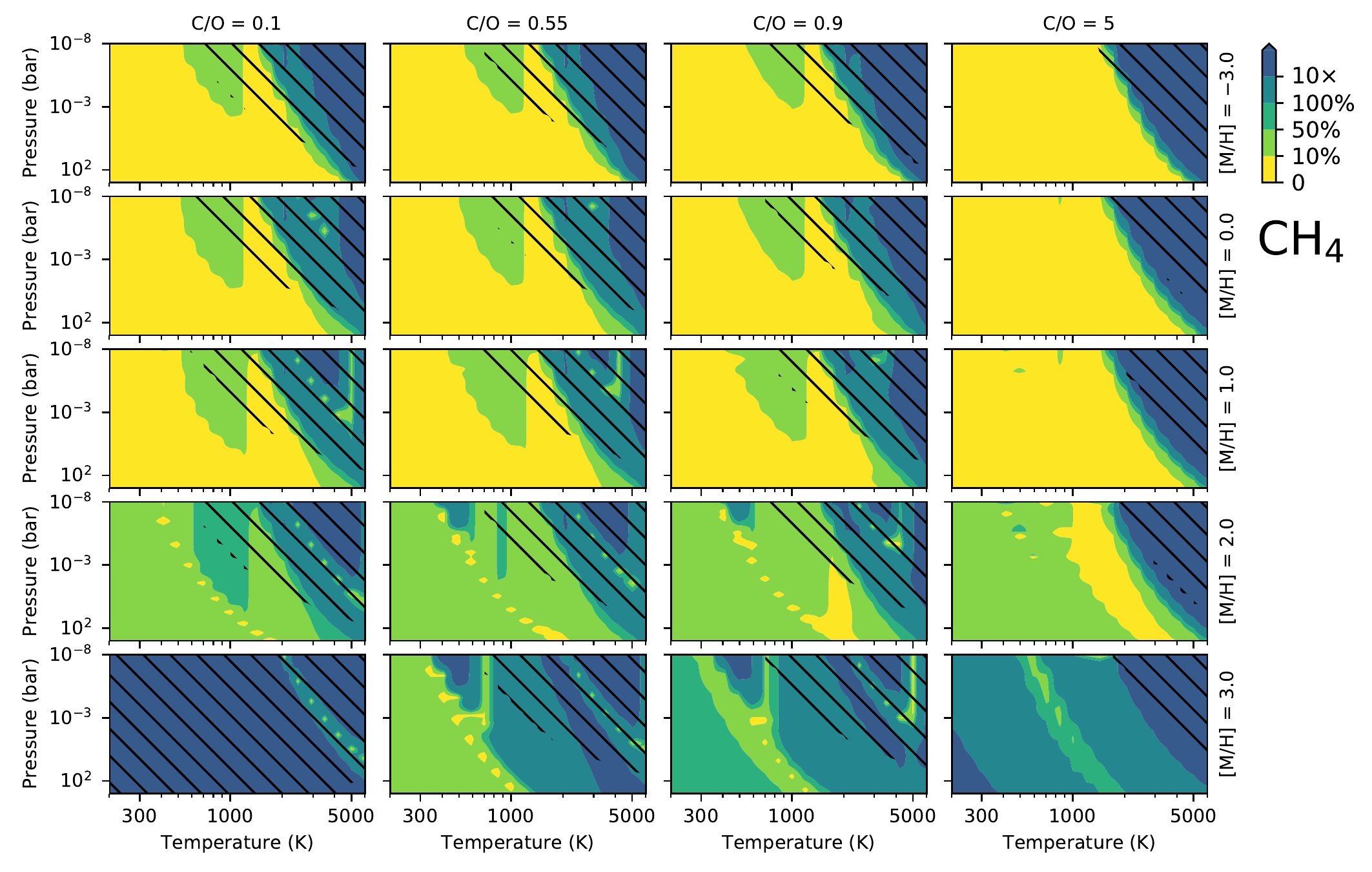}
\includegraphics[width=\linewidth, clip]{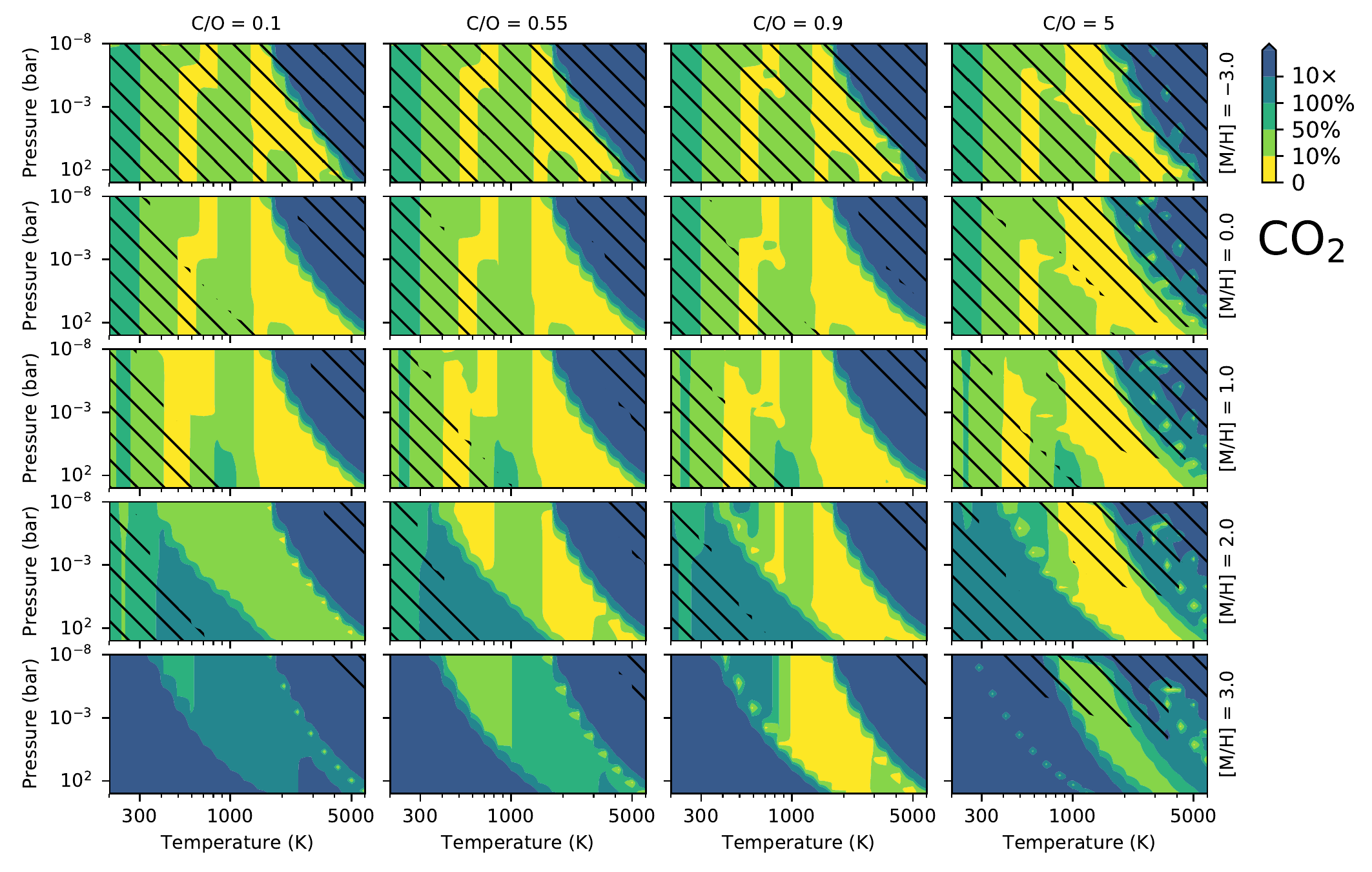}
\caption{Difference between {\TEA} and {\rate} abundances for {\methane}
(top panels) and {\carbdiox} (bottom panels).  See description in
caption of Fig.~\ref{fig:bench1}.}
\label{fig:bench2}
\end{figure*}

\begin{figure*}[p]
\centering
\includegraphics[width=\linewidth, clip]{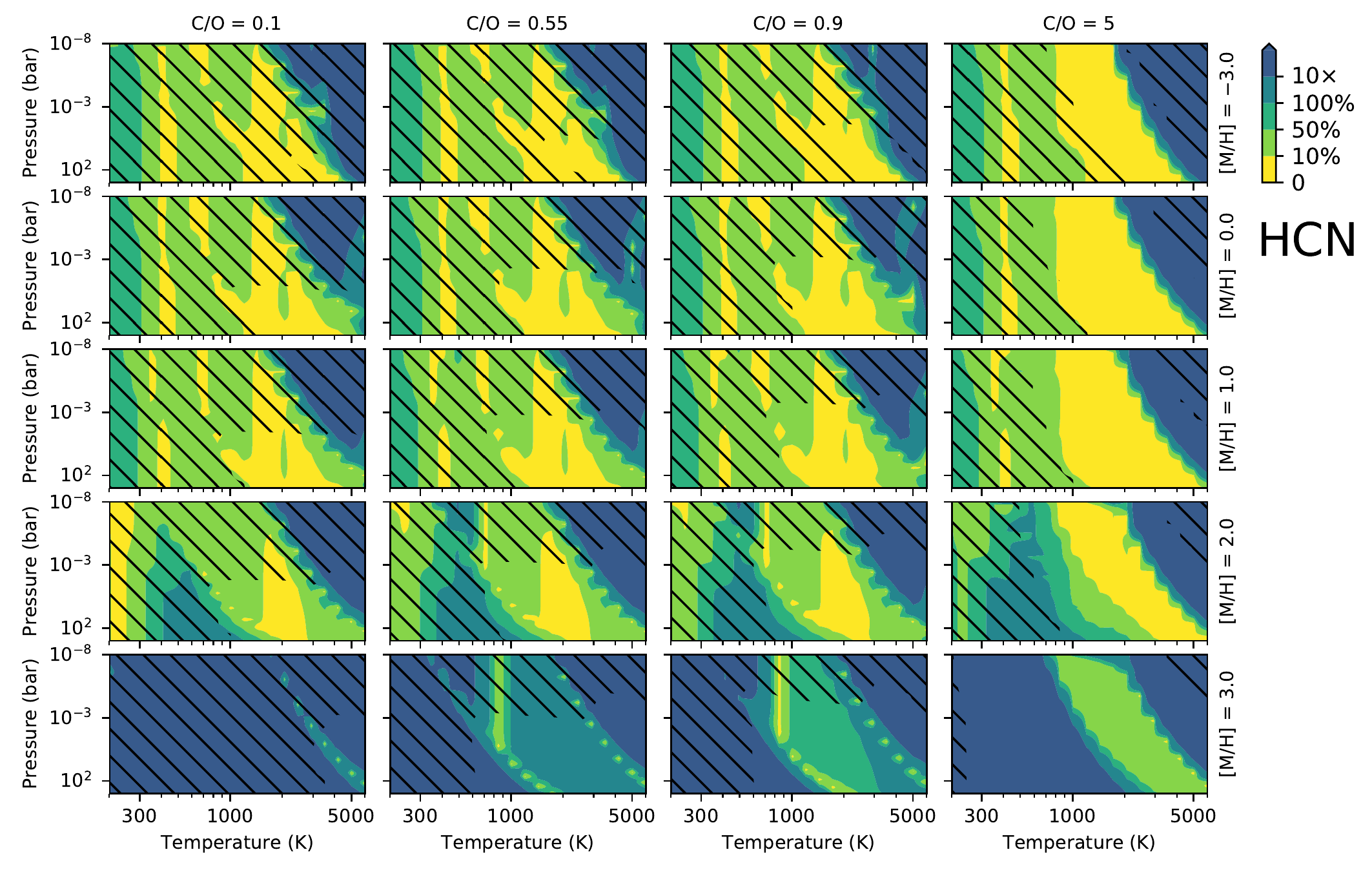}
\includegraphics[width=\linewidth, clip]{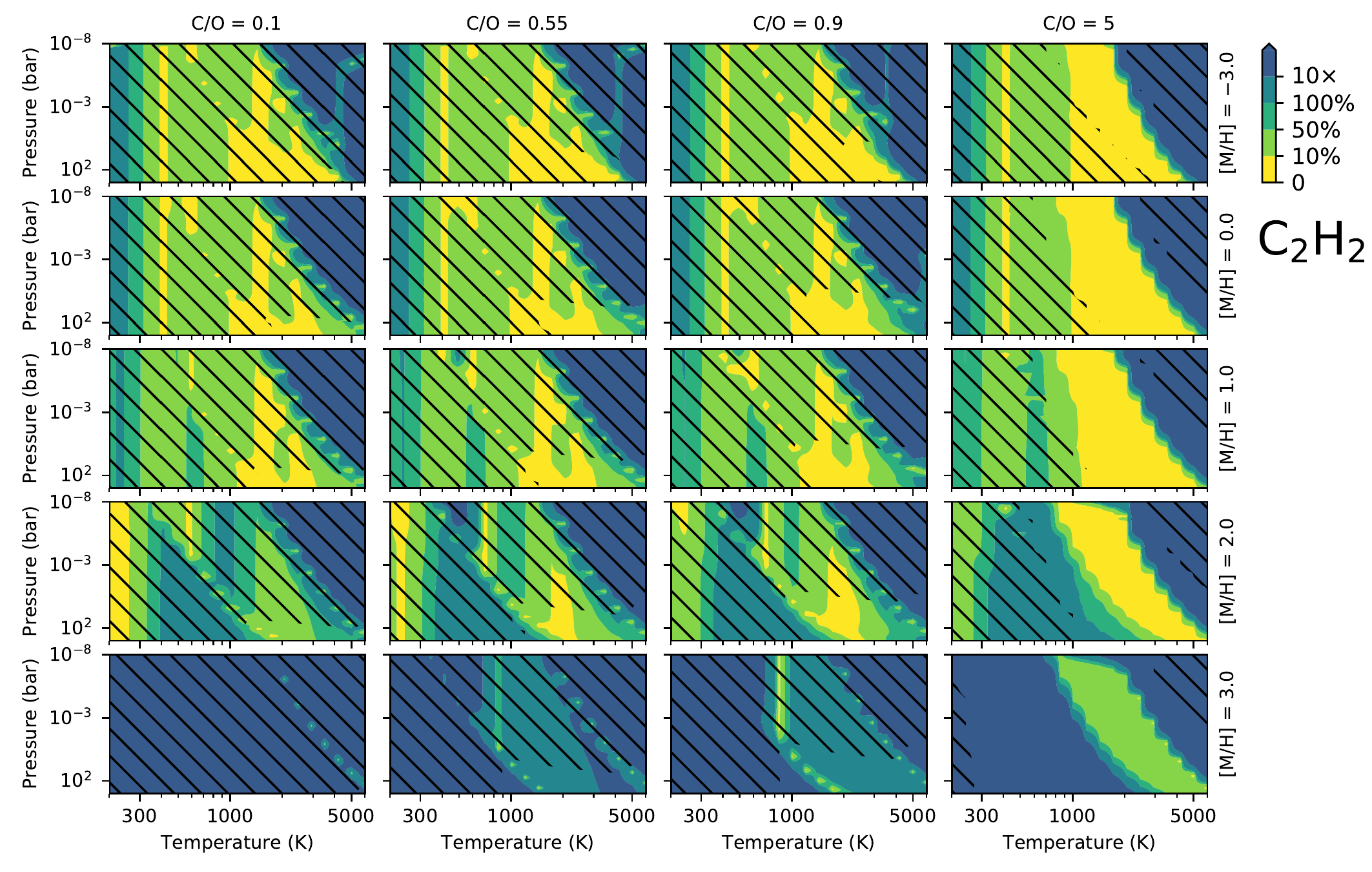}
\caption{Difference between {\TEA} and {\rate} abundances for HCN (top panels)
and {\acetylene} (bottom panels).  See description in caption of
Fig.~\ref{fig:bench1}.}
\label{fig:bench3}
\end{figure*}

\begin{figure*}[p]
\centering
\includegraphics[width=\linewidth, clip]{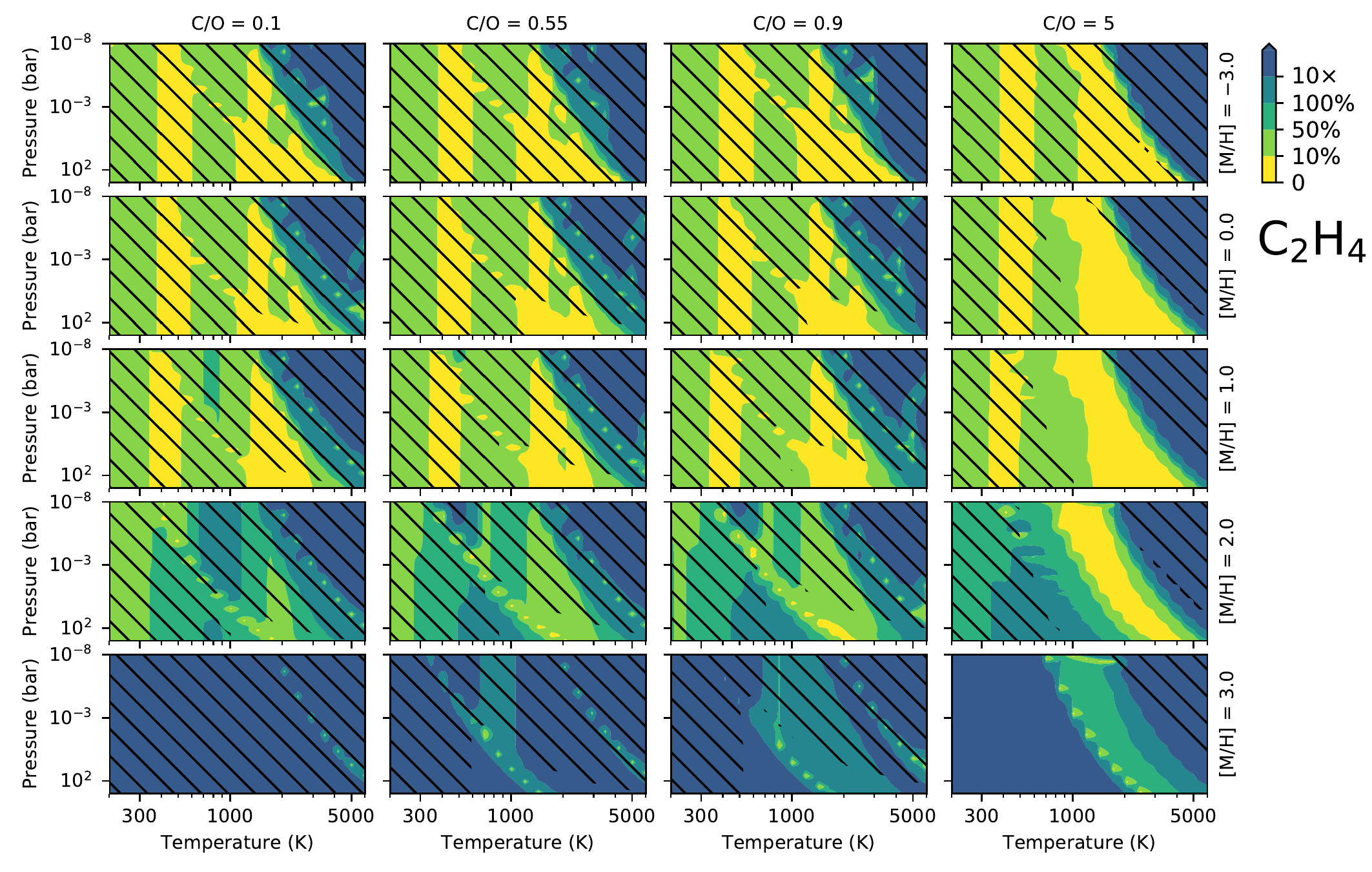}
\includegraphics[width=\linewidth, clip]{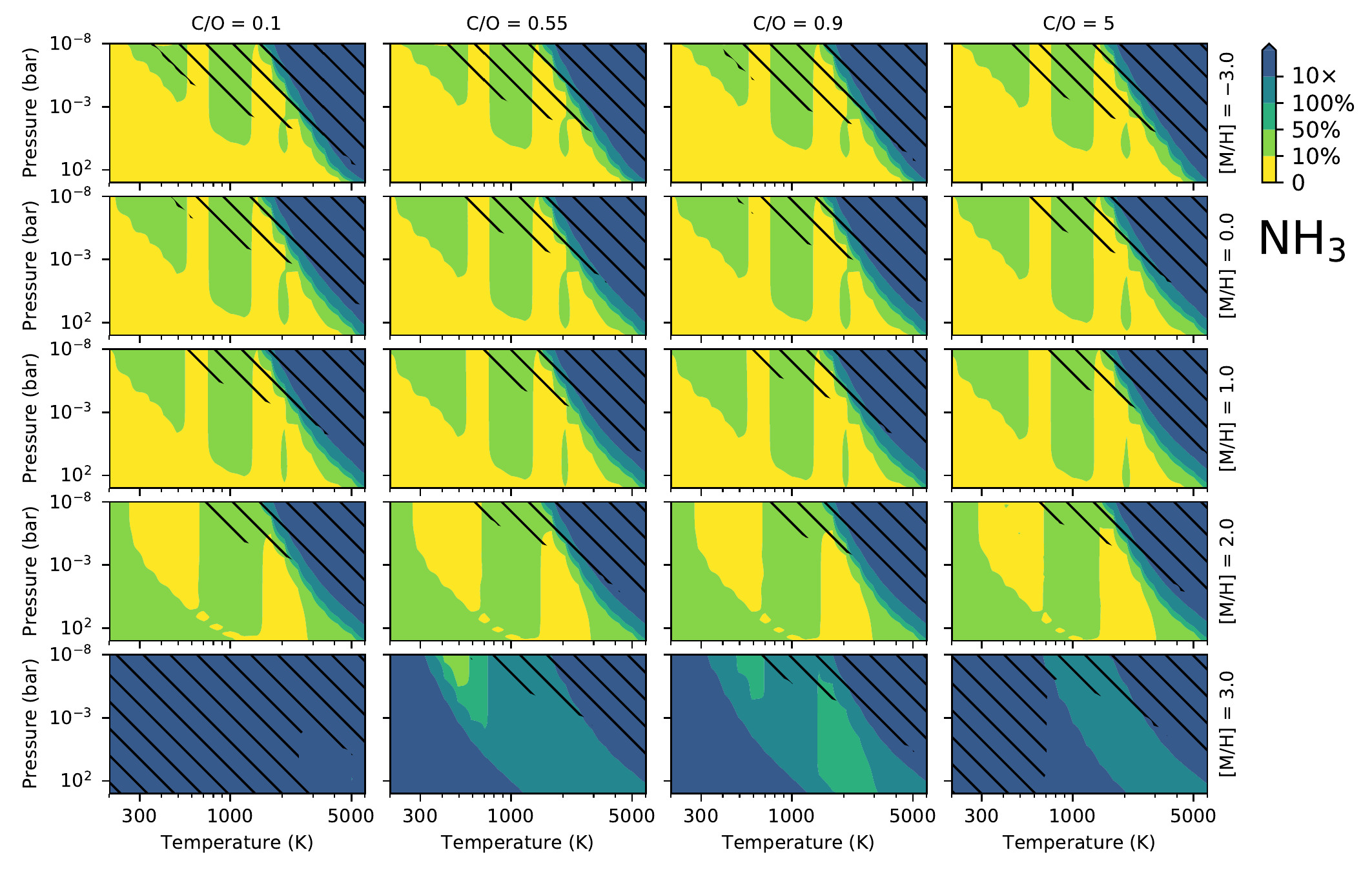}
\caption{Difference between {\TEA} and {\rate} abundances for {\ethylene}
(top panels) and {\ammonia} (bottom panels).  See description in
caption of Fig.~\ref{fig:bench1}.}
\label{fig:bench4}
\end{figure*}

\begin{figure*}[p]
\centering
\includegraphics[width=\linewidth, clip]{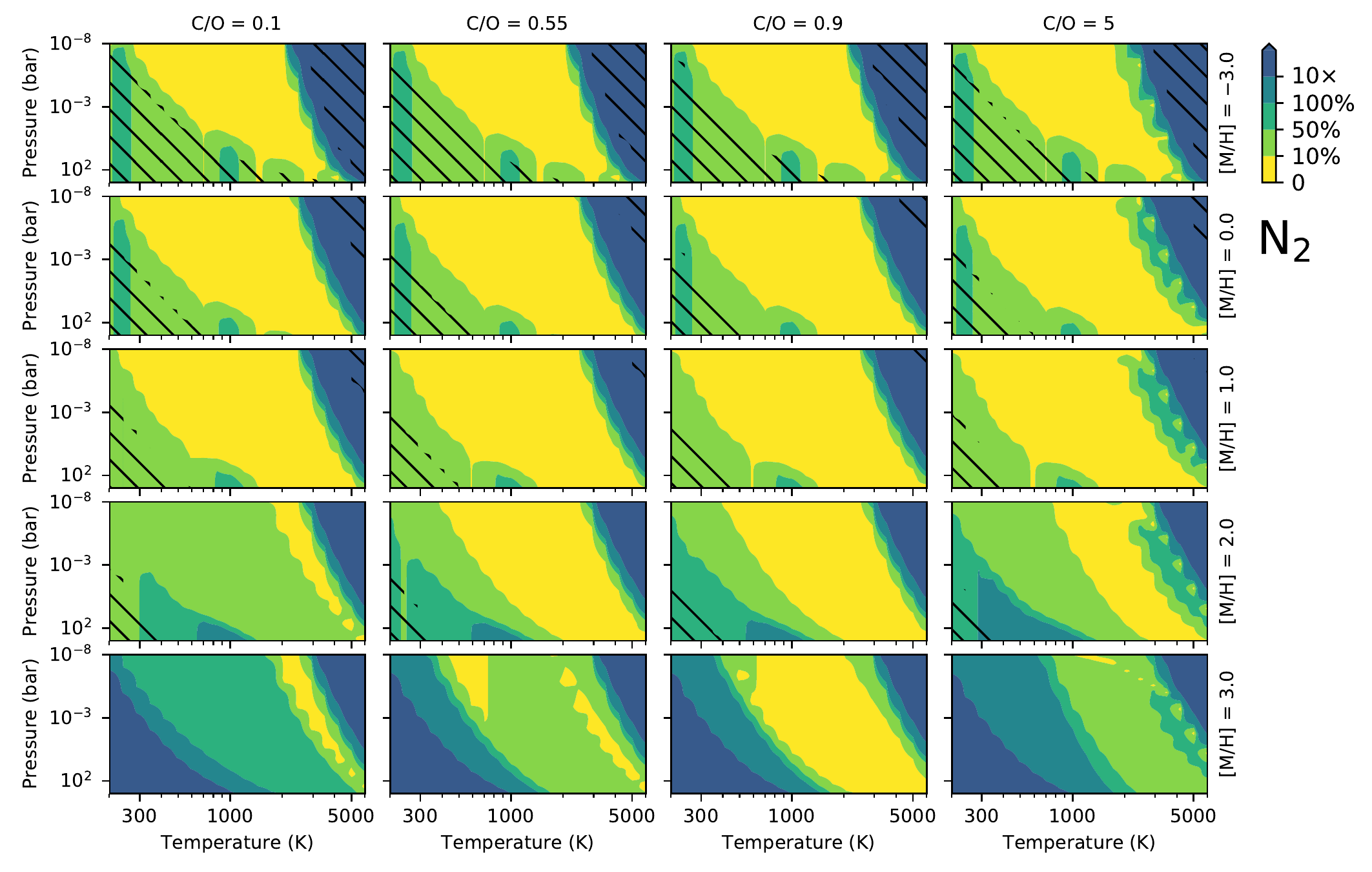}
\includegraphics[width=\linewidth, clip]{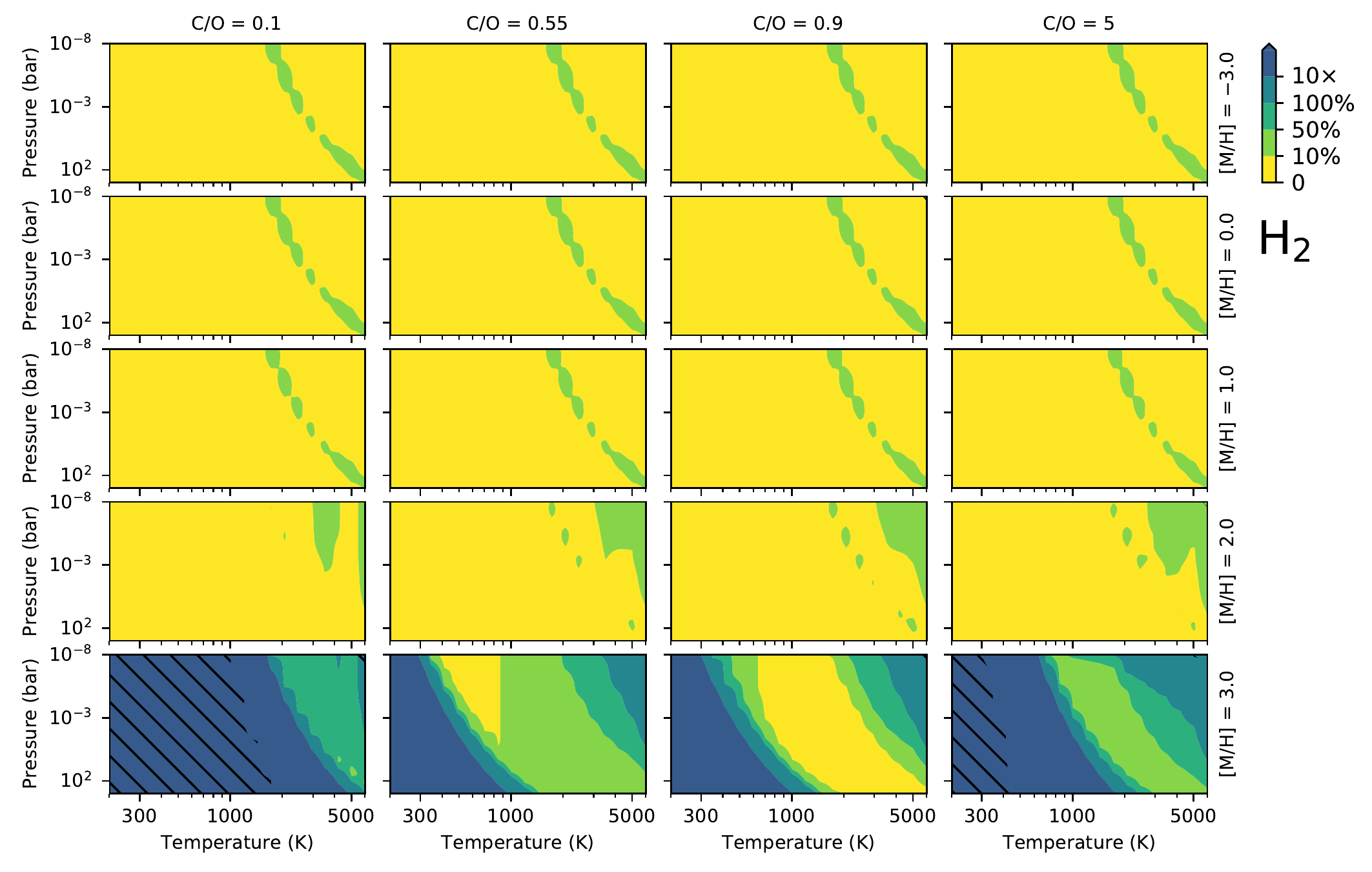}
\caption{Difference between {\TEA} and {\rate} abundances for {\nitrogen}
(top panels) and {\molhyd} (bottom panels).  See description in
caption of Fig.~\ref{fig:bench1}.}
\label{fig:benc5}
\end{figure*}

\begin{figure*}[tb]
\centering
\includegraphics[width=\linewidth, clip]{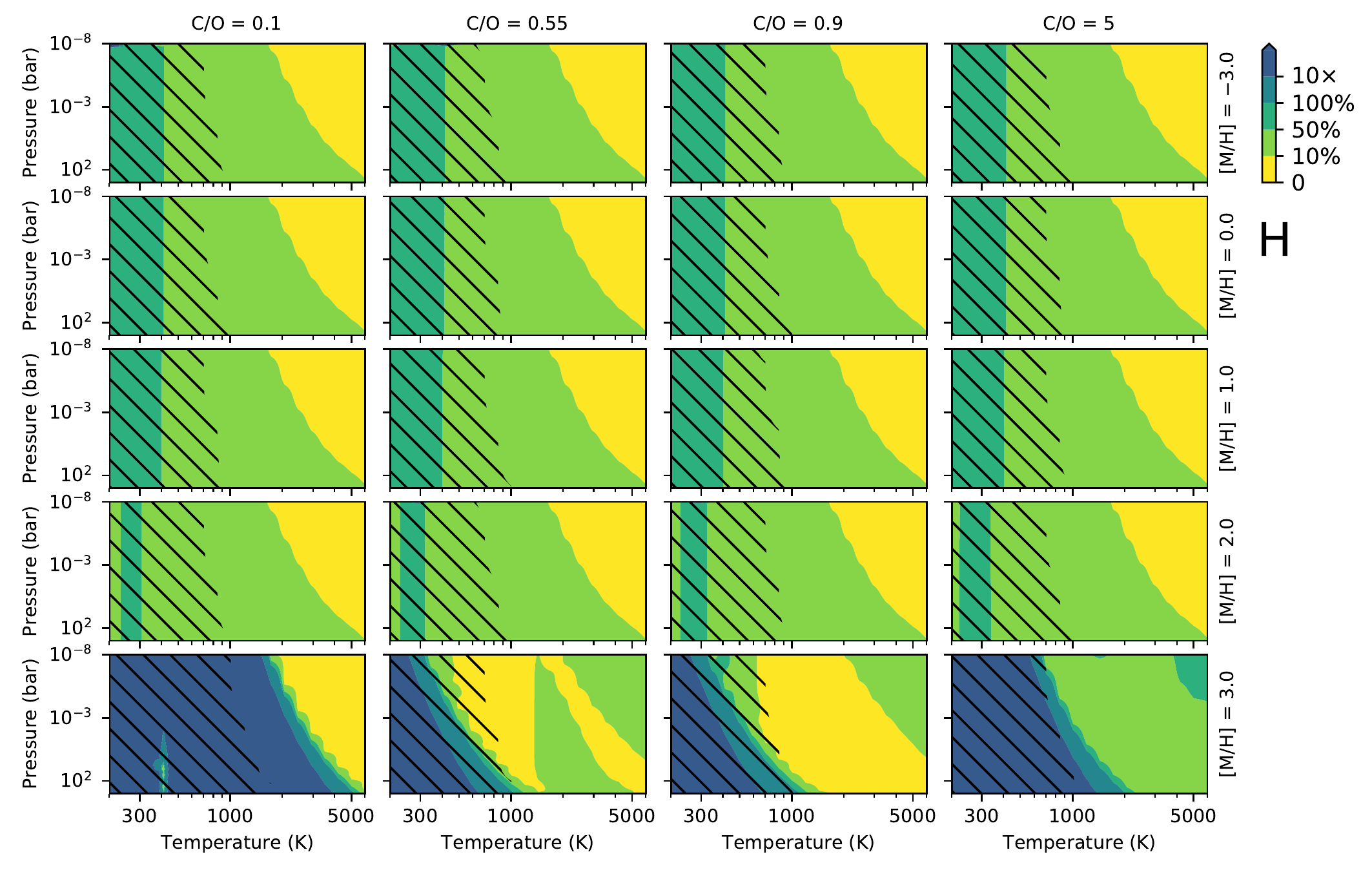}
\caption{Difference between {\TEA} and {\rate} abundances for H.  See
description in caption of Fig.~\ref{fig:bench1}.}
\label{fig:bench6}
\end{figure*}

Figures \ref{fig:bench1}--\ref{fig:bench6} show the accuracy of
{\rate} for each species.  For each panel, we compute the metal
elemental abundances such that they obey the labeled metallicity:
\begin{equation}
 {\rm [M/H]} = \log\frac{(\ncarbon+\nnitrogen+\noxygen)\phantom{\sb{\odot}}}
                         {(\ncarbon+\nnitrogen+\noxygen)\sb{\odot}},
\end{equation}
combined with the labeled elemental ratio for C/O =
$\ncarbon/\nnitrogen$, while maintaining
a fixed solar elemental ratio for C/N =
$(\ncarbon/\nnitrogen)\sb{\odot}\approx 4$.

In general, we find a good agreement between {\rate} and {\TEA}, with
no major variations in the accuracy as a function of C/O ratios.
The accuracy of {\rate} roughly correlates with the species
abundances, meaning that the code performs better for the species that 
are more relevant for spectroscopy, particularly for the main species
that determine the infrared spectrum of sub-stellar objects (e.g.,
{\water}, CO, and {\methane}).
For the relatively abundant species (mixing fractions larger than
$\ttt{-10}$), the typical accuracy is better than 10\%; for the less
abundant species, the typical accuracy is better than 50\%.  The
accuracy for each species varies differently with temperature,
pressure, and metallicity, and is largely proportionally to the species
abundance.

\begin{figure*}[th]
\centering
\includegraphics[width=\linewidth, clip, trim=0 110 20 0]{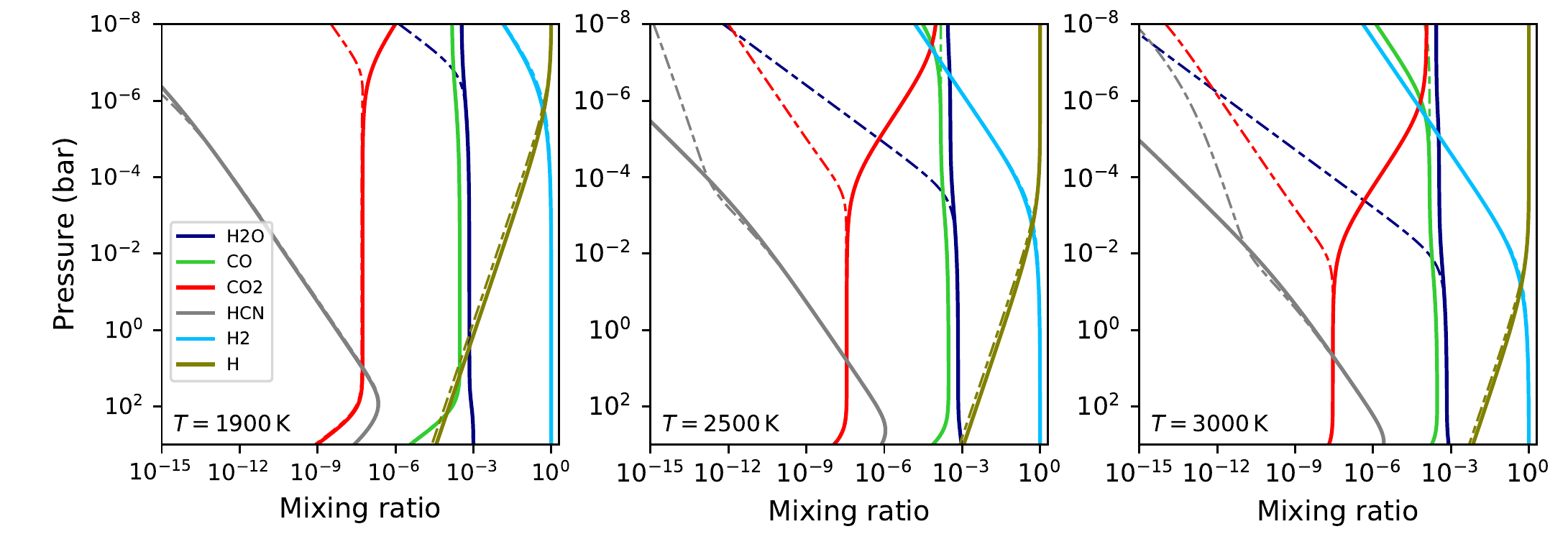}
\caption{
Comparison between analytic (solid curves) and {\TEA} (dashed curves)
thermochemical-equilibrium abundances, at increasing temperatures
(left, middle, and right panels).  Only a selected subset of species
id shown (see legend). The atmosphere has a nearly solar composition
with $\ncarbon=1.5\tttt{-4}$, $\noxygen=5.0\tttt{-4}$, and
$\nnitrogen=7.0\tttt{-5}$.  As the temperature increases, the analytic
values deviate from the thermochemical-equilibrium values at
progressively lower pressures, due to the dissociation of the
molecular species.}
\label{fig:hot}
\end{figure*}

We find two distinctive regions where {\rate} does not perform well.
There is a sharp decrease in accuracy at a combination of high
temperatures and low pressures.  Starting at $T \gtrsim 2000$~K at the
{$\sim$\microbar} level, the boundary of this region propagates toward
higher pressures as the temperature increases (see top-right corner of
panels in Figs.~\ref{fig:bench1}--\ref{fig:bench6}).  In this case,
the molecules start to dissociate into their atomic form, which we do
not account for metals (i.e., C, N, and O).  In particular, the
analytic approach overestimates the abundances for {\water} and
{\carbdiox}, compared to their expected thermochemical-equilibrium
values (Figure~\ref{fig:hot}).  We emphasize that this is not an
implementation issue, the outputs reflect exactly what the equations
indicate.  Rather, the analytic approach fails to fully trace the
physics of the problem.
Similarly, the accuracy starts to decrease proportionally with
the increasing metallicity for super solar values.  From 100 to 1000
times solar metallicity, {\rate} produces abundances deviating by a
factor of 2 to more than 10 times from the expected values,
respectively.  In this case, the assumption of hydrogen-dominated
atmosphere is no longer valid, and the analytic results become
inaccurate.

\section{Conclusions}
\label{sec:conclusions}

We have developed a framework that expands on the work
of \citet{HengEtal2016apjAnalyticCO}, \citet{HengLyons2016apjAnalyticHCO},
and \citet{HengTsai2016apjAnalyticHCNO}, to analytically compute
thermochemical-equilibrium abundances for a chemical system with
arbitrary temperature, pressure, and elemental abundances.  This
approach computes the mole mixing ratios for {\water}, {\methane}, CO,
{\carbdiox}, {\ammonia}, {\acetylene}, {\ethylene}, HCN, {\nitrogen},
{\molhyd}, atomic H, and He, by finding the roots of a univariate
polynomial expression.  We implemented this approach into the {\rate}
open-source package (compatible with Python 2.7 and 3), which is
available at \href{https://github.com/pcubillos/rate}
{https://github.com/pcubillos/rate}.

To obtain the analytic equations, we developed a general and nearly
effortless approach to find the polynomial coefficients using
{\sympy}, which facilitates future development of this approach.  In
our system of equations, we accounted for atomic hydrogen (in addition
to molecular hydrogen), which is the dominant hydrogen-bearing
molecule at high temperatures and low pressures, due to dissociation.
We treat He as a constant-abundance non-interacting species.

We found three key factors that improve the numerical stability of the
analytic approach over prior efforts: we apply a more reliable
algorithm to solve for the polynomial roots; we identify and avoid the
regimes where solving polynomials for {\water} or CO is numerically
unstable; and we identify the regimes where one can neglect HCN, and
thus decouple the nitrogen chemistry from the carbon and oxygen
equations to find simpler polynomial expressions.

This code computes reliable abundances over a range of pressures from
$\ttt{-8}$ to $\ttt{3}$~bar; temperatures from 200
to $\sim$2000~K; and elemental metallicities from
$\ttt{-3}$ to $\sim\!\ttt{2}\times$ solar
values.
Beyond these boundaries, the assumptions of this framework do
not apply anymore, and thus, the the analytic solutions become less
accurate, i.e., molecules dissociate into atoms at higher temperatures,
and the composition is no longer hydrogen-dominated at higher metallicities.

Our framework produces abundances with an accuracy better than 10\% of
their expected values for the more abundant species (mixing ratios
larger than $\ttt{-10}$), and therefore, those more relevant for
spectroscopy.  Ultimately, different applications will have different
accuracy requirements.  For example, for exoplanet atmospheric
characterization, we do not expect to find abundance uncertainties
much better than 50\% for the dominant
species \citep{GreeneEtal2016apjJWSTtransitCharacterization}.  Our
improvements enable us to use the fast analytic approach to compute
reliable equilibrium abundances over a wide range of temperatures,
pressures, and elemental abundances.  With the exception of the
extreme ultra-hot Jupiters, this framework is of particular interest
to model and retrieve the observed atmosphere of most giant-exoplanet
atmospheres, since the molecules considered here are the main species
that dominate the infrared and optical spectrum of these planets.

The Reproducible Research Compendium of this article is available at
\dataset[10.5281/zenodo.2529532]{\doi{10.5281/zenodo.2529532}}.

\acknowledgments

We thank the anonymous referee for his/her time and valuable comments.
We thank contributors to the Python Programming
Language (see Software below); the free and open-source community. This research
has made use of NASA's Astrophysics Data System Bibliographic
Services. J.B. is supported by NASA through the NASA
ROSES-2016/Exoplanets Research Program, grant NNX17AC03G.  We drafted
this article using the aastex6.2 latex
template \citep{AASteamHendrickson2018aastex62}, with some further style
modifications that are available
at \href{https://github.com/pcubillos/ApJtemplate}
{https://github.com/pcubillos/ApJtemplate}.

\software{
{\rate} (\href{https://github.com/pcubillos/rate}
              {https://github.com/pcubillos/rate}),
\textsc{vulcan}\footnote{\href{http://github.com/exoclime/VULCAN}
                              {http://github.com/exoclime/VULCAN}}
\citep{TsaiEtal2017apjsVULCAN},
{\TEA}\footnote{\href{https://github.com/dzesmin/TEA}
                     {https://github.com/dzesmin/TEA}}
\citep{BlecicEtal2016apsjTEA},
\textsc{Numpy} \citep{vanderWaltEtal2011numpy},
\textsc{SciPy}  \citep{JonesEtal2001scipy},
{\sympy} \citep{MeurerEtal2017pjcsSYMPY},
\textsc{Matplotlib} \citep{Hunter2007ieeeMatplotlib},
\textsc{IPython} \citep{PerezGranger2007cseIPython},
AASTeX6.2\footnote{\href{https://doi.org/10.5281/zenodo.1209290}
                        {https://doi.org/10.5281/zenodo.1209290}}
\citep{AASteamHendrickson2018aastex62},
ApJtemplate (\href{https://github.com/pcubillos/ApJtemplate}
                  {https://github.com/pcubillos/ApJtemplate}),
and \textsc{BIBMANAGER}\footnote{
\href{http://pcubillos.github.io/bibmanager}
     {http://pcubillos.github.io/bibmanager}}
\citep{Cubillos2019bibmanager}.
}






\end{document}